\documentclass[10pt,letterpaper,twocolumn,aps,prl]{revtex4}
\usepackage{ae}
\usepackage{aecompl}
\usepackage[T1]{fontenc}
\usepackage[latin1]{inputenc}
\usepackage{graphicx}

\makeatletter

\usepackage{graphicx}
\begin{document}

\title{Directional tunneling escape from nearly spherical optical resonators}

\author{Scott Lacey, Hailin Wang, David H. Foster and Jens U. N\"{o}ckel}

\email{noeckel@uoregon.edu}

\homepage{http://oco.uoregon.edu}

\affiliation{Department of Physics, University of Oregon, Eugene, OR 97403}

\date{07/17/2003, Phys. Rev. Lett. \textbf{91}, 033902 (2003)}

\begin{abstract}
We report the surprising observation of directional tunneling escape
from nearly spherical fused silica optical resonators, in which most
of the phase space is filled with non-chaotic regular trajectories.
Experimental and theoretical studies of the dependence of the far-field
emission pattern on both the degree of deformation and the excitation
condition show that non-perturbative phase space structures in the
internal ray dynamics profoundly affect tunneling leakage of the whispering
gallery modes. 
\end{abstract}


\maketitle
Electromagnetic fields in a uniform dielectric sphere can be calculated
in much the same way as quantum mechanical wave functions in a spherically
symmetric potential. The sphere exhibits whispering-gallery (WG) modes,
which are long-lived resonances with electromagnetic fields confined
near the surface \cite{Nussenzveig}. For small deviations from the
spherical shape, $\epsilon\equiv(R_{max}-R_{min})/(R_{max}+R_{min})\ll1$
(where $R_{max},\, R_{min}$ are the maximal and minimal radii), perturbative
treatments \cite{LaiHM} are routinely employed to infer deformation
parameters from the splitting of azimuthal degeneracy of the WG modes
\cite{Seguin}. Certain strongly non-spherical resonators, on the
other hand, can be analyzed with methods from quantum chaos such as
random-matrix theory or periodic-orbit expansions \cite{Stockmann99}.
Many generic resonator shapes, however, fall into a transition regime
in which none of these known methods apply globally. When entering
this regime from the perturbative side, calculations may encounter
singularities and undefined limits \cite{BerryMVPhysTod}. Experimental
studies in this regime can thus provide insight into how nature resolves
the competition between perturbative and non-perturbative physics,
here with the resonator shape as a control parameter.

In this paper we present studies of far-field emission patterns and
resonance lifetimes of deformed fused-silica microspheres. The experimental
and theoretical studies of the dependence of the far-field emission
pattern on both the degree of deformation and the excitation condition
show that highly directional tunneling escape can occur in microspheres
with small deformation ($\epsilon\approx1\%$) and with a large size
parameter ($kR=2\pi R/\lambda\approx785$, where $\lambda\approx800$nm,
$R\approx100\mu$m). These results are completely unexpected in the
ray optics limit or in the earlier perturbative wave treatment with
$\epsilon$ as a small parameter. The observed emission pattern shows
that on one hand, the ray model breaks down in predicting the escape
mechanism of WG modes for resonators with small deformation. On the
other hand, the nonperturbative phase space structures predicted by
the ray model can profoundly affect tunneling leakage of the WG modes.
Even at extremely small deformation, nonperturbative phase space structures
still persist and are directly responsible for directional tunneling
escape. This intricate interplay between ray and wave dynamics provides
essential physical insights into properties of weakly deformed WG
resonators, especially, for those in which the phase space is filled
with non-chaotic regular trajectories. Directional emission patterns
have previously been observed in more strongly deformed resonators
where a significant fraction of the internal rays shows chaotic motion
\cite{ChangCSN00}.

Deformed microspheres were formed by melting together two spheres
of similar sizes. The individual spheres were fabricated by melting
an optical fiber tip with a focused $\mathrm{CO}_{2}$ laser beam.
The spheres were brought into contact and heated until surface tension
produced a completely convex surface. By carefully controlling the
temperature of the glass using the $\mathrm{CO}_{2}$ laser, it was
possible to repeatedly reduce the degree of deformation. For reference,
we define the elongated axis of the resulting prolate spheroid, which
also connects the centers of the two original spheres, to be the $x$-axis.
The $z$-axis is defined by the remaining fiber stem, which breaks
the rotational symmetry about the $x$-axis, making the deformed microsphere
completely non-axisymmetric. The stem was held by a fiber chuck for
easy manipulation of the sphere. Images of these deformed microspheres
taken from three orthogonal directions have been shown in an earlier
study \cite{LaceyW01}.

\begin{figure}
\begin{center}\includegraphics[%
  width=0.90\columnwidth]{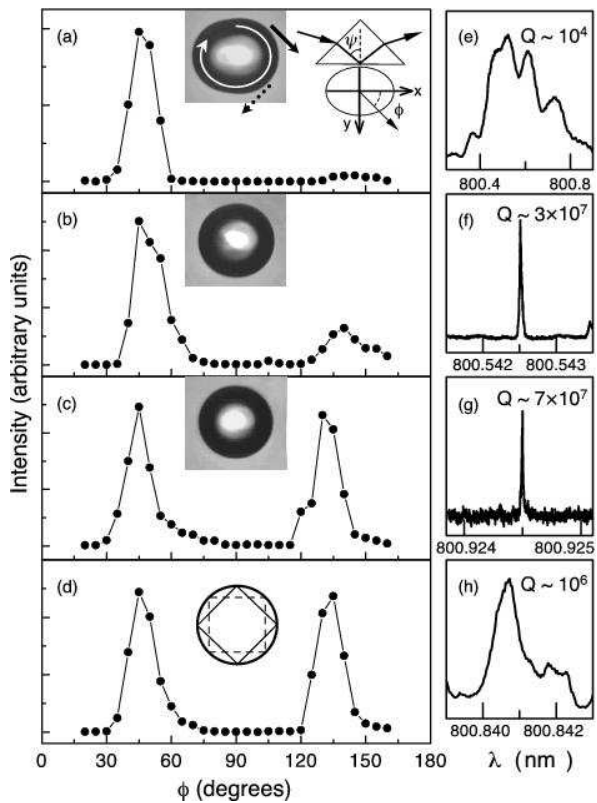}\end{center}

\caption{\label{fig:directionexperiment}(a)-(d): Far-field emission patterns
of WG modes. Insets: bottom view of the resonators showing the progression
of shapes in the $x$-$y$ cross-section, from which we determine
\protect$\epsilon=6.7\%$ (a), \protect$\epsilon=2.2\%$ (b), and
\protect$\epsilon=1.2\%$ (c,d). WG modes were launched at \protect$\sin\chi_{0}\approx1$
in (a)-(c), and at \protect$\sin\chi_{0}\approx0.8$ in (d); $\chi$
is the internal angle of incidence with respect to the surface normal.
(e)-(h): the spectra corresponding to the modes in (a)-(d), from which
we deduce the Q-factors. Inset to (d): stable (solid) and unstable
(dashed) 4-bounce orbits in a quadrupole at \protect$\epsilon=1.2\%$.}
\end{figure}

To investigate emission properties and resonance lifetimes of the
deformed microspheres, we use frustrated total internal reflection
at the surface of a prism with refractive index $n=1.7$ to launch
individual traveling-wave WG modes near the $x$-$y$ plane of the
spheroid (Fig. \ref{fig:directionexperiment}a inset) \cite{newfocus}.
The initial angle of incidence $\chi_{0}$ with which the WG modes
are launched in the microsphere can be controlled by adjusting the
angle of incidence $\psi$ in the prism.

WG mode spectra were obtained by measuring the far-field emission
intensity as a function of the excitation wavelength. Since the stem
holding the resonator is a strong leakage pathway, the observed long-lived
WG modes must have internal field patterns that do not overlap with
the stem region. This is made possible by the fact that the resonator
is slightly flattened in the $z$ direction (along the stem axis).
Experiments as well as ray simulations show that this deformation
stabilizes the rays in the vicinity of the $x$-$y$ plane. Our directionality
measurements correspondingly were performed in this plane. By scanning
the detector while keeping the resonator fixed, we recorded WG mode
spectra at various angles, $\phi$ (from the $x$-axis), to construct
the far-field emission pattern at a given microsphere deformation
$\epsilon$. The microsphere was then reheated to reduce $\epsilon$
and the far-field emission pattern was measured again, leading to
the evolution shown in Figure \ref{fig:directionexperiment}. The
excitation beam is s-polarized, and the detection scheme is polarization-insensitive.
The far-field emission patterns are independent of the polarization
of the excitation laser beam. Spheroids with similar $\epsilon$ exhibit
qualitatively the same behavior in terms of mode spectra, Q-factors,
and emission patterns.

Beginning with a strongly deformed microsphere, the emission was observed
to have a strong peak at $\phi=45^{\circ}$ in the far field {[}see
Fig. \ref{fig:directionexperiment} (a){]}. Since clockwise traveling
waves were excited, as indicated by the white arrow in the sphere
image in Fig. \ref{fig:directionexperiment} (a), this far-field emission
direction corresponds to light escaping tangential from the region
at $\phi=-45^{\circ}$ on the surface of the microsphere, shown by
the solid arrow in the sphere image. A much smaller peak ($5\%$ of
the large peak height) in the far-field emission pattern was observed
at $\phi=135^{\circ}$, corresponding to light escaping from the region
at $\phi=45^{\circ}$ on the microsphere surface (dotted arrow). As
the deformation was reduced, the far-field emission peak at $\phi=135^{\circ}$
grew to about one quarter of the $\phi=45^{\circ}$-peak height in
Fig. \ref{fig:directionexperiment} (b). In Fig. \ref{fig:directionexperiment}
(c), the two peaks reached nearly equal intensity. The bright emission
spots at $\phi=\pm45^{\circ}$ on the resonator surface can be viewed
directly with a CCD camera. For reference, we call the pattern in
Fig. \ref{fig:directionexperiment} (a) \emph{asymmetric} and the
pattern in Fig. \ref{fig:directionexperiment} (c) \emph{symmetric}
(around $\phi=90^{\circ}$). Figures \ref{fig:directionexperiment}
(e)-(g) also show that as the deformation was reduced, the Q-factor
of the relevant WG modes increased by nearly four orders of magnitude.

The measurements discussed thus far were performed with input condition
$\chi_{0}\approx90^{\circ}$ ($\sin\chi_{0}\approx1$). Figure \ref{fig:directionexperiment}
(d) shows the far-field emission pattern from the same microsphere
as Fig. \ref{fig:directionexperiment} (c) but with light launched
at $\sin\chi_{0}\approx0.8$. The emission pattern is nearly identical
to Fig \ref{fig:directionexperiment} (c) where $\sin\chi_{0}\approx1$,
although the corresponding Q-factor is nearly two orders of magnitude
smaller {[}Figs. \ref{fig:directionexperiment} (g) and (h){]}.

The observed emission patterns in Fig. \ref{fig:directionexperiment}
(a)-(d) \emph{disagree} with the intuitive expectation that WG modes
in an oval resonator should preferentially emit tangential to the
points of highest curvature, into the far-field direction $\phi=\pm90^{\circ}$.
This is what one obtains when modeling our spheroids as triaxial \emph{ellipsoids},
whose internal ray dynamics exhibits no chaos \cite{WaalkensWD99},
independently of the axis ratios.

The observed emission pattern of the most deformed microsphere, Fig.
\ref{fig:directionexperiment} (a), is well explained by a ray model.
The peak at $\phi=45^{\circ}$ can be attributed to an effect known
as dynamical eclipsing \cite{ChangCSN00}. In the ray model, a WG
mode corresponds to light rays trapped close to the perimeter of the
dielectric resonator by total internal reflection, which prevents
light escape unless a \emph{critical angle} $\chi_{c}\equiv\arcsin1/n=43.6^{\circ}$
is reached, where $n=1.45$ is the refractive index of the fused silica.
At small but finite $\epsilon$, any oval can be approximated by the
first terms of a multipole expansion, which after proper choice of
origin is \emph{quadrupolar}. In this limit, a stable 4-bounce orbit
shaped like a diamond forms with its sharp vertices at the highest-curvature
points of the resonator with an angle of incidence $\chi_{4}\approx45^{\circ}$
{[}inset to Fig. \ref{fig:directionexperiment} (d){]}. This creates
{}``islands of stability'' in phase space \cite{Stockmann99}: rays
launched near the diamond orbit will retain similar reflection points
and angles even if they do not close onto themselves after four bounces.
At the deformation used in Fig. \ref{fig:directionexperiment} (a),
most of the phase space supporting the WG mode in question is chaotic.
However, chaotic rays cannot penetrate the 4-bounce islands; this
means chaotic WG rays are prevented from refractively escaping at
the points of highest curvature, because $\chi_{4}\approx\chi_{c}$.
Instead, as trajectories flow around the islands in one direction,
refractive escape occurs near $\phi=-45^{\circ}$ on the resonator
surface. From this circulation around the islands, \emph{asymmetric}
far-field emission patterns result as a hallmark of refractive escape.
This is what we observe in Fig. \ref{fig:directionexperiment} (a).

This ray mechanism was originally proposed and tested for effectively
2D systems where islands of stability and chaotic regions are mutually
exclusive. In our 3D spheroids, the same diamond-shaped stable orbit
exists in or near the $x$-$y$ plane, and refractively escaping rays
in its vicinity have lifetimes that correspond to $\leq10$ round
trips in the resonator, translating to a Q-factor near $10^{4}$.
During this time, a given ray can be considered as moving in a cross-sectional
plane that may be inclined and slowly precessing around the $z$-axis.
The variation of cross-sectional shape, $\Delta\epsilon$, probed
by such rays causes no significant differences in the size of the
4-bounce islands \cite{ChangRC99}. Therefore, rays in the 3D resonator
behave as in a corresponding fixed 2D resonator during a relatively
short time preceding an escape event, and in particular show dynamical
eclipsing.

However, \emph{refractive} escape ceases to be the dominant leakage
mechanism at the very small deformation used in Figs. \ref{fig:directionexperiment}
(c) and (d), since in this case chaotic diffusion is no longer effective
and the stable diamond orbit itself also becomes fully confined by
total internal reflection. Those classical rays that do escape refractively,
moreover, are not able to produce the \emph{symmetric} emission patterns
observed in Figs. \ref{fig:directionexperiment} (c) and (d) \cite{Foot1}.
Hence, the emission process in these cases is governed by evanescent
escape, i.e. a tunnel coupling between internal and radiation fields.

Tunneling is the only decay mechanism in an ideal sphere where each
WG mode in the plane of excitation corresponds to a unique azimuthal
angular momentum number $m$ with respect to the $z$-axis, which
is semiclassically related to $\chi_{0}$ by $\sin\chi_{0}=m/(nkR)$,
using the fact that the modes of interest remain close to the $x$-$y$
plane and hence have total angular momentum $l\approx m$. At $kR\gg1$,
the tunneling escape rate is negligible when $\sin\chi_{0}\approx1$
but accelerates exponentially as $\sin\chi_{0}$ approaches $\sin\chi_{c}$.
In a deformed microsphere, the angle of incidence $\chi$ varies as
a well-defined function of $\phi$ provided it covers an angular momentum
range where chaos is neglibible \cite{NockelS97}. A circulating ray
with varying $\chi(\phi)>\chi_{c}$ will then escape with exponentially
strong selectivity near the \emph{minima} of $\chi(\phi)$.

The emission locations and directions observed in the experiment indicate
that the minima of $\chi(\phi)$ lie near the corners of the unstable
rectangular 4-bounce orbit {[}inset to Fig. \ref{fig:directionexperiment}
(d){]}. We determined from ray tracing that this only occurs near
but above $\chi_{c}$. Thus, the symmetric far-field patterns in Figs.
\ref{fig:directionexperiment} (c) and (d) are a \emph{tunneling probe}
for the minima of the \emph{ray-optical} $\chi(\phi)$ in the vicinity
of the critical angle $\chi_{c}$. To further corroborate this interpretation,
recall that the emission patterns as shown in Figs. \ref{fig:directionexperiment}
(c) and (d) are insensitive to the initial angle of incidence, $\chi_{0}$,
as they should be if the detected light originates near $\sin\chi\approx\sin\chi_{c}$.
The question remains how the light coupled into the resonator is able
to reduce its angular momentum from a high value corresponding to
$\sin\chi_{0}$ to the much lower $\sin\chi_{c}$. Our observation
shows that such a dynamical process is present in the spheroid, but
does not unambiguously reveal its mechanism. A possible mechanism
is Arnol'd diffusion, a phase-space transport process that crucially
depends on the fact that the resonators are in fact three-dimensional
\cite{LaceyW01}, with no axial symmetry. Even without resolving this
question, we can nevertheless conclude that the phase space region
near $\sin\chi_{c}$ is responsible for the observed emission patterns.

\begin{figure}
\begin{center}\includegraphics[%
  width=0.90\columnwidth]{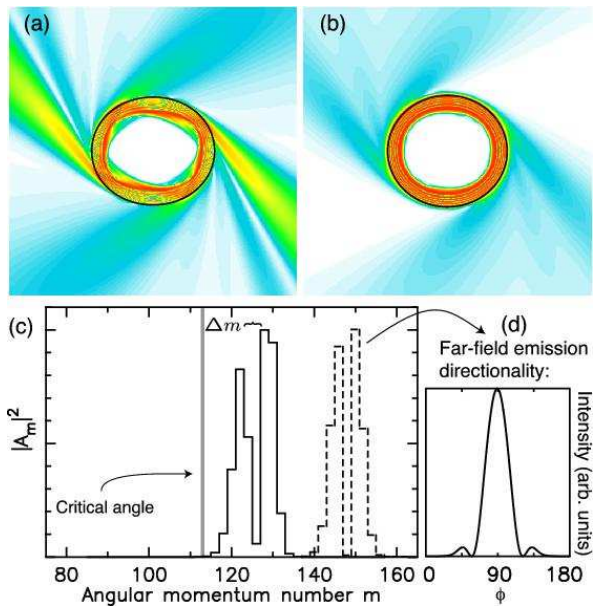}\end{center}

\caption{\label{fig:waveplots}Top: Intensity pattern of WG modes with parameters
\protect$kR=112.063$, \protect$\epsilon=6.5\%$ (a) and \protect$kR=112.452$,
\protect$\epsilon=3.4\%$ (b). The Q-factors are \protect$4\times10^{4}$
(a) and \protect$2\times10^{5}$ (b), and the overall directionality
agrees with that in Fig. \ref{fig:directionexperiment} (b) and (d),
respectively. The values of \protect$\epsilon$ differ from Fig.
\ref{fig:directionexperiment} because the simulation approximates
the unknown details of the experimental shape by a pure quadrupole
\cite{NockelS97}. Bottom: solid histogram in (c) shows the angular-momentum
distribution of the mode in (b). A very high-Q WG mode at the same
deformation (c, dashed), with \protect$kR=113.229$ and \protect$Q=3\times10^{14}$
radiates predominantly into \protect$\phi=90^{\circ}$, as shown
in the far-field pattern (d). }
\end{figure}

Our initial assumption of a 2D model for the emission process, justified
by the approximate 2D nature of the 3D ray trajectories for intermediate
times as discussed earlier, is therefore consistent with the fact
that no long-time properties of the internal ray dynamics are needed
to explain the emission directions. With the same justification, we
also performed numerical wave calculations by modeling the shape felt
by modes in the $x$-$y$ plane as the cross section of a quadrupolar
cylinder. Figures \ref{fig:waveplots} (a) and (b) plot the intensity
patterns of two WG modes under traveling-wave excitation \cite{NockelS97,BraunILNSSVWW00}.
The calculation assumes s-polarized light and uses size parameters
near $kR\approx113$ \cite{numericnote}.

The emission patterns in Figs. \ref{fig:waveplots} (a) and (b) agree
qualitatively with the asymmetric and the symmetric emission patterns
shown in Figs. \ref{fig:directionexperiment} (b) and (d), respectively.
To understand the physical mechanism for the emission in Fig. \ref{fig:waveplots}
(b), we plot in Fig. \ref{fig:waveplots} (c) the distribution of
angular momentum numbers for the WG mode in (b). Figure \ref{fig:waveplots}
(c) shows negligible overlap with the window for refractive escape
below the critical angle $\chi_{c}$ (corresponding to a \emph{critical
angular momentum} $m_{c}=kR\approx113$). This indicates that the
emission from this mode is due to \emph{tunneling} escape. By comparison,
both refractive and tunneling escape contribute to the emission in
Fig. \ref{fig:waveplots} (a), with the refractive escape playing
a dominant role.

The peak splitting of width $\Delta m\approx4$ in Fig. \ref{fig:waveplots}
(c) is a straightforward consequence of the oval deformation: as the
wave circulates around the resonator with varying radius between $R_{min}$
and $R_{max}$, its angular momentum oscillates but has high probabitity
of being near its extrema, given by the extrema of $m\approx r(\phi)nkR\sin\chi(\phi)$
over the surface. The fact that the minima of $\chi(\phi)$ do \emph{not}
occur at $\phi=0^{\circ},\,180^{\circ}$ on the surface is due to
the 4-bounce island structure in the vicinity of $\sin\chi=\sin45^{\circ}$.

For comparison, Fig. \ref{fig:waveplots} (c) also shows the distribution
of angular momentum number of an ultra-high Q WG mode with emission
directionality as expected for the ellipse {[}see Fig. \ref{fig:waveplots}
(d){]}. The mode is confined at a high $m$, corresponding to $\sin\chi_{0}\approx0.91$.
Note that here the tunneling loss is negligible compared with scattering
or absorption loss of the material that limits the actual Q-factor
of fused silica microspheres to of order $10^{9}$ \cite{Gorodetsky}.
The two modes in Fig. \ref{fig:waveplots} (c) have practically no
overlap in angular momentum space. The qualitative difference in the
far-field patterns of these two modes further confirms that the mode
shown in Fig. \ref{fig:directionexperiment} (c), which was excited
at $\sin\chi_{0}\approx1$, is not confined to $\sin\chi_{0}\approx1$.

In essence, our experiment exploits the peculiar feature that for
fused silica the critical angle lies near the phase-space islands
corresponding to the stable 4-bounce orbit. These islands are the
dominant structure in the WG region and persist even at small $\epsilon$
where most of the phase space is filled with non-chaotic, {}``regular''
trajectories. The island formation is a \emph{non-perturbative} consequence
of the breakdown of conservation laws \cite{WaalkensWD99}, which
qualitatively distinguishes the generic quadrupolar shape from an
ideal ellipse even though the two shapes differ only to second order
in $\epsilon$. The emission characteristics of a silica spheroid
in the seemingly trivial small-$\epsilon$ regime will \emph{continue}
to be strongly affected by ray patterns that wrap around the perimeter
in approximately four bounces.

Whether any non-perturbative structure in the ray dynamics can be
\emph{resolved} by the wave field, depends on $kR$ \cite{PrangeNZ99}:
a directionality measurement will be able to distinguish the peaks
at $\phi=90^{\circ}\pm45^{\circ}$ if the conjugate angular momentum
$m$ satisfies the uncertainty relation $\Delta\phi\,\Delta m\geq1/2$
with $\Delta\phi\approx\pi/4$. This implies $\Delta m>2/\pi\approx1$,
which in our spheroids translates to a fluctuating angle of incidence
of $\Delta\sin\chi=\Delta m/(nkR)\approx10^{-3}$. Even the least
deformed resonator ($\epsilon\approx1\%$) exceeded this estimated
resolution threshold significantly.

It is truly remarkable that this intricate interplay between ray and
wave dynamics for which chaos plays no dominant role, can be exploited
to engineer WG modes that can feature both high-Q and highly directional
emission. This makes WG resonators with \emph{small} deformation highly
promising for a variety of applications \cite{Kippenberg,Armani},
such as microlasers, nonlinear optics, and quantum information processing.

This work was supported in part by NSF under grants No. DMR9733230
and No. DMR0201784.


\begin{thebibliography}{10}
\bibitem{newfocus}The measurement was performed with a tunable diode laser (New Focus).
\bibitem{Foot1}Even at small deformation, the separatrix region we are probing is
created by the homoclinic tangle of the unstable, rectangle-shaped
periodic orbit shown dashed in the inset to Fig. 1 (d), and this tangle
exhibits an asymmetry which will produce asymmetric emission whenever
the escape is predominantly ray-based.
\bibitem{numericnote}We have performed the same calculations at wave numbers $kR\approx70$,
and close to the current limits of our numerical technique at $kR\approx200$.
All these results are consistent with the ones presented in Fig.~2,
showing that the phase-space structure causing the observed emission
is already fully resolved in Fig.~2. Extrapolating to the experiment
(Fig. 1) at $kR\approx785$, we expect no further changes within the
angular resolution of the detector apparatus.
\end{thebibliography}
\end{document}